# Enhancement of Spin-Charge Conversion Efficiency for $Co_3Sn_2S_2$ across Transition from Paramagnetic to Ferromagnetic Phase

Takeshi Seki,[1*] Yong-Chang Lau,[1,2] Junya Ikeda,[1] Kohei Fujiwara,[1] Akihiro Ozawa,[1] Satoshi Iihama,[3,4] Kentaro Nomura,[1,5] and Atsushi Tsukazaki[1,6]

[1] *Institute for Materials Research, Tohoku University, Sendai 980-8577, Japan*

[2] *Institute of Physics, Chinese Academy of Sciences, Beijing 100190, China*

[3] *Frontier Research Institute for Interdisciplinary Sciences, Tohoku University, Sendai 980-8578, Japan*

[4] *WPI Advanced Institute for Materials Research, Tohoku University, Sendai 980-8577, Japan*

[5] *Department of Physics, Kyushu University, Fukuoka 819-0395, Japan*

[6] *Center for Science and Innovation in Spintronics, Core Research Cluster, Tohoku University, Sendai 980-8577, Japan*

\* e-mail: takeshi.seki@tohoku.ac.jp

**Abstract:**

Co$_3$Sn$_2$S$_2$ (CSS) is one of the shandite compounds and becomes a magnetic Weyl semimetal candidate below the ferromagnetic phase transition temperature ($T_C$). In this paper, we investigate the temperature ($T$) dependence of conversion between charge current and spin current for the CSS thin film by measuring the spin-torque ferromagnetic resonance (ST-FMR) for the trilayer consisting of CSS / Cu / CoFeB. Above $T_C \sim 170$ K, the CSS / Cu / CoFeB trilayer exhibits the clear ST-FMR signal coming from the spin Hall effect in the paramagnetic CSS and the anisotropic magnetoresistance (AMR) of CoFeB. Below $T_C$, on the other hand, it is found that the ST-FMR signal involves the dc voltages ($V_{dc}$) not only through the AMR but also through the giant magnetoresistance (GMR). Thus, the resistance changes coming from both AMR and GMR should be taken into account to correctly understand the characteristic field angular dependence of $V_{dc}$. The spin Hall torque generated from the ferromagnetic CSS, which possesses the same symmetry as that for spin Hall effect, dominantly acts on the magnetization of CoFeB. A definite increase in the spin-charge conversion efficiency ($\xi$) is observed at $T < T_C$, indicating that the phase transition to the ferromagnetic CSS promotes the highly efficient spin-charge conversion. In addition, our theoretical calculation shows the increase in spin Hall conductivity with the emergence of magnetic moment at $T < T_C$, which is consistent with the experimental observation. (236 words)

# I. Introduction

Creation, manipulation and detection of the spin angular momentum flow, called spin current ($J_s$), are vital to an efficient operation of the spintronic devices. High efficiency in the conversion process between charge current ($J_c$) and $J_s$ is required for improving the device performance, reducing the power consumption, and leading to multi-functionalities. One of the promising ways for spin-charge conversion is to exploit the spin Hall effect (SHE) [1,2], which is mostly studied in nonmagnets (NMs). The conversion from $J_c$ to $J_s$ via SHE is expressed as

$$\mathbf{j}_s = \left(\frac{\hbar}{2e}\right)\alpha_{\mathrm{SH}}(\mathbf{s}\times\mathbf{j}_c), \qquad (1)$$

where $\mathbf{j}_c$ and $\mathbf{j}_s$ are charge current density and spin current density, respectively, $\alpha_{\mathrm{SH}}$ is the spin Hall angle, $e$ (< 0) is the electric charge of an electron, $\hbar$ is the reduced Planck constant, and $\mathbf{s}$ is the quantization axis of electron spin. $\alpha_{\mathrm{SH}}$ is a key parameter related to the spin-orbit coupling and the efficiency of spin-charge conversion. Thus, many studies were devoted to the development of spin Hall material with large $\alpha_{\mathrm{SH}}$. It has been demonstrated that not only a NM but also a ferromagnet (FM) is utilized for the spin-charge conversion [3-27]. In the early stage, the conversion from $J_s$ to $J_c$ in FM was reported [3-5]. Then, it was predicted that the anomalous Hall effect (AHE) also generates $J_s$, which is called spin anomalous Hall effect (SAHE) [6], and the SAHE was examined for several FMs experimentally and theoretically [12,17,24,27]. Apart from the SAHE, other spin-charge conversion processes allow to generate $J_s$ with a variety of spin polarization vectors such as spin precession process [8,9]. Stimulated by these theoretical predictions, many experimental works have recently reported the generation of $J_s$ in the FM bulks or at the FM interfaces [15].

For developing a spin Hall material, it is a guide to exploit the topological features in an electronic band structure [28,29]. $Co_3Sn_2S_2$ (CSS) is one of the shandite compounds, and has attracted much attention as a promising candidate for magnetic Weyl semimetals [30-36]. Thanks to its specific band structure, CSS exhibits half-metallicity and huge anomalous Hall effect (AHE) below the ferromagnetic phase transition temperature ($T_C$) that was reported to be ~175 K in bulk [37-40]. One may expect that the ferromagnetic CSS shows the highly efficient spin-charge conversion. The paramagnetic CSS above $T_C$ is also an attractive candidate for improvement of the spin-charge conversion efficiency by carefully considering its band structure. Our group has already demonstrated that the Fermi-level tuning of the paramagnetic CSS by elemental doping is very effective to obtain the enhanced spin Hall conductivity at room temperature [41]. Although our previous study that suggests the paramagnetic CSS family is a promising spin Hall material, no one has understood yet how the ferromagnetic phase transition accompanying with the variation of electronic band affects the spin-charge conversion process in the case of the CSS. For example, it is unclear if the spin Hall effect for the CSS is enhanced in the vicinity of $T_C$ as shown in the previous studies for other ferromagnets [42,43]. This is an essential question to understand the spin-charge conversion process in the magnetic topological materials.

In this paper, the temperature ($T$) dependence of spin-charge conversion in the CSS thin film is investigated by measuring the spin-torque ferromagnetic resonance (ST-FMR) for the trilayer consisting of CSS / Cu / $Co_{20}Fe_{60}B_{20}$ (CFB). By carefully analyzing the field angular dependence of ST-FMR signal, we examine the influence of ferromagnetic phase transition on the spin-charge conversion efficiency, and discuss the major process of spin-charge conversion in the ferromagnetic CSS. In addition, we study the spin-charge conversion process of CSS based on the effective tight-binding model. This paper is composed of the following sections. **Section II** describes the experimental procedures including the film preparation and the details of ST-FMR measurement method. **Section III** is devoted to the experimental results and discussion, and is divided into five subsections: temperature dependence of Hall effect and magnetoresistance, spin torque ferromagnetic resonance spectra, quantitative analysis of spin torque, enhanced spin-charge conversion efficiency for ferromagnetic CSS, and discussion. In **Sec. IV**, theoretical calculation is shown. Finally, the conclusion is given in **Sec. V**.

## II. Experimental Procedure

The 15 nm-thick CSS layer was prepared on the $Al_2O_3$ (0001) substrate by radio-frequency magnetron sputtering. A CSS layer and a Si-O capping layer were deposited at a substrate temperature of 400 °C under an Ar gas of 0.5 Pa. Subsequently, in situ annealing was performed at 800 °C in a vacuum to promote the crystallization of CSS. The Si-O capping layer was used to prevent re-evaporation during the annealing. The *c*-axis oriented growth of CSS was confirmed by x-ray diffraction measurement, which was reported in Ref. [41]. The CSS / Si-O film was set into the Ar ion milling chamber to remove the Si-O capping layer. During this ion milling process, the surface of CSS layer was also etched, and the designed CSS layer thickness was reduced to 9.8 nm. After the removal of Si-O layer, subsequently Cu (1.8 nm)/CFB (2 nm)/Al-O (5 nm) were deposited at room temperature by employing the ion beam sputtering system.

For the ST-FMR measurement, the CSS / Cu / CFB / Al-O stack was patterned into a rectangular-shaped element with 10 μm width and 40 μm length, and the Au electrodes of coplanar waveguide were fabricated through the use of photolithography and Ar ion milling. The ST-FMR measurement was carried out with the setup employed in the previous studies [12,17,44,45]. As illustrated in **Fig. 1(a),** the radiofrequency (rf) current ($I_{rf}$) was applied along the *x* direction, and the in-plane angle of external magnetic field (*H*) was set at $\theta$ from the *x* direction. The rf power of 15 dBm was applied from a signal generator [**Fig. 1(b)**], inducing an oscillating transverse magnetic field in the $y$ direction. The excitation frequency ($f$) of $I_{rf}$ was varied in the range from 6 GHz to 16 GHz. The device resistance [$R(t)$] oscillated through the anisotropic magnetoresistance (AMR) and/or giant magnetoresistance (GMR) effect at the condition that $H$ matched the resonance field ($H_{Res}$). As a result, applied $I_{rf}(t)$ [$= I \cos(2\pi ft)$] and oscillating $R(t)$ [$\propto \cos(2\pi ft)$] generated a rectification dc voltage ($V_{dc}$), which was detected by a lock-in amplifier. In addition to the coplanar waveguide device, the Hall devices were prepared for measuring the temperature dependence of the longitudinal resistance ($R_{xx}$) and the transverse resistance ($R_{yx}$). The coplanar waveguide device and the Hall device were fabricated on the identical substrate. The ST-FMR spectra for the coplanar waveguide device were measured employing the temperature-variable rf-compatible probe station. The maximum in-plane |$H$| of the probe station was 6 kOe. The values of $R_{xx}$ and $R_{yx}$ for the Hall device were measured with the superconducting magnet allowing to increase $H$ up to 70 kOe along in-plane and out-of-plane directions.

The magnetic properties of the blanket films were measured using a superconducting quantum interference device magnetometer.

## III. Experimental Results and Discussion

### A. Temperature dependence of Hall effect and magnetoresistance

**Figure 2(a)** shows $R_{yx}$ versus out-of-plane magnetic field ($H_z$) for the Hall device measured at $T$ = 300 K, 200 K, 150 K, and 100 K. The linear variation without hysteresis in the of $R_{yx}$ - $H_z$ curves is observed at $T$ = 300 K and 200 K. These small changes in $R_{yx}$ originate from the AHE of CFB layer. The CFB layer is in-plane magnetized at $H_z$ = 0 Oe and becomes out-of-plane magnetized as $H_z$ is increased up to 10 kOe. When $T$ was reduced to 150 K, the AHE of CSS layer contributes to $R_{yx}$. The further reduction of $T$ to 100 K leads to the clear hysteretic behavior and the large remanent $R_{yx}$ at $H_z$ = 0 Oe because of the out-of-plane spontaneous magnetization of CSS. These results indicate that $T_C$ of the present CSS layer exists between 150 K and 200 K.

Employing the identical device, $R_{xx}$ versus in-plane $H$ was measured. This corresponds to the magnetoresistance (MR) curve measurement. **Figure 2(b)** displays the MR curve measured at $T$ = 300 K, where the red (blue) curve denotes the MR curve under the in-plane $H$ sweeping from positive (negative) to negative (positive). A small but sharp change in $R_{xx}$ at low $H$ and the following gradual $R_{xx}$ variation are observed as $H$ is increased, which come from the AMR and the forced effect, respectively, of the CFB layer. When $T$ is reduced down to 50 K, the large resistance change is observed at low $H$ [**Fig. 2(c)**]. The inset of **Fig. 2(c)** shows the corresponding MR curve enlarged at the low $H$ region. This is attributable to the GMR effect due to the change in the relative configuration of in-plane magnetization components between CSS and CFB although the CSS is mostly magnetized along the out-of-plane direction, which will be explained later. One of the remarkable features for the MR curves at $T < T_C$ is that the exchange-biased-like behavior, *i.e.* an asymmetric MR curve with respective to the zero magnetic field is observed if the applied in-plane $H$ is insufficient for fully saturating the in-plane component of CSS. This is because the CSS layer possesses the high magnetic anisotropy and the resultant large switching field. **Figures 2(d) and 2(e)** correspond to the minor MR curves at $T$ = 50 K measured with the narrow $H$ sweep ($\pm$ 20 kOe), where the in-plane $H$ = + 20 kOe and – 20 kOe, respectively, were applied during the device cooling from $T$ = 300 K to 50 K. In **Fig. 2(d)** with + 20 kOe-field cooling, the positive exchange-bias-like field, *i.e.* the sharp resistance change only in the negative $H$ region (see the inset) is observed whereas the negative exchange-bias-like field is induced for the case of **Fig. 2(e)** with – 20 kOe-field cooling. As seen in the AHE hysteresis of **Fig. 2(a)**, the CSS is mostly magnetized along the out-of-plane direction. Nevertheless, the GMR effect accompanied by the exchange-biased-like behavior appears when the in-plane $H$ was applied. Considering these facts, the in-plane magnetized CSS exists near the interface, and a part of in-plane magnetized CSS shows soft magnetic behavior and the other shows hard magnetic behavior. The appearance of in-plane magnetization component may be due to the CSS damaged during the Ar ion milling process. In addition, the results of minor MR curves [**Figs. 2(d) and 2(e)**] indicate that the direction of magnetic field cooling determines the initially magnetized direction for the hard magnetic CSS. The magnetic moments in the soft magnetic CSS are easily switched by $H$, but are coupled to the hard magnetic phase, resulting in the exchange-bias-like behavior. **Figure 2(f)** schematically illustrates possible magnetic structures in CFB and CSS. The coexistence of in-plane magnetized soft and hard magnetic CSS is one scenario to explain the observed MR curves. One may think that the exchange-bias-like behavior is due to the appearance of antiferromagnetic (AFM) phase [46] or geometric frustration intrinsic to the kagome network of magnetic ions [47]. According to the paper reporting the AFM phase [46], the AFM phase appears in the limited $T$ region around $T_C$. **Figure 2(g)** displays the $T$ dependence of $R_{xx}$ and $R_{yx}$ and **Fig. 2(h)** displays the $T$ dependence of $\Delta R$ defined as the difference between $R_{xx}$ at $H = + 20$ kOe and $- 20$ kOe in the minor MR curve [see **Fig. 2(e)**]. From these temperature dependences, we point out the following two results. First, the $T$ dependence of $R_{yx}$ suggests that $T_C$ for the present CSS is ~ 170 K. Second, $\Delta R$ is observed from around 100 K to even at 10 K. These facts indicate that the exchange-bias-like mechanism is maintained at $T$ much lower than the limited $T$ region of AFM reported in Ref. 46, and we consider that the AFM phase is not responsible for the exchange-bias-like behavior observed in this study. On the contrary, it was reported that the geometric frustration appears at $T < T_G = 125$ K at which a spin glass phase is considered to be formed [47]. As shown in **Fig. 2(h)**, $\Delta R$ becomes remarkable at $T$ below $T_C$. This result may suggest the contribution of geometric frustration. In order to explain the shape of MR curve, however, the phase pinning the soft magnetic CSS should possess the spontaneous magnetization. Considering this point, we currently think that not the geometric frustration but the hard magnetic phase of CSS gives rise to the exchange-biased-like MR shift for the present samples. The coplanar waveguide devices for the ST-FMR measurement exhibited the $T$ dependence of MR effect similar to those observed for the Hall devices, which is given in **Appendix 1**.

**B. Spin torque ferromagnetic resonance spectra**

**Figures 3(a) and 3(b)** display $V_{dc}$ as a function of $H$ measured at $T = 300$ K and 80 K, respectively, by varying the in-plane field angle of $\theta$. $f$ was fixed at 16 GHz. The results for $T = 300$ K show clear ST-FMR at $\theta \neq 0º$, 90º, 180º, and 270º. The single-peak spectral shapes indicate that the resonance spectra are composed of ferromagnetic resonance for the single ferromagnet. Since the CSS is paramagnetic at $T = 300$ K, the ST-FMR originates from the magnetization dynamics of the CFB layer induced by spin-torque. When $T$ is reduced to 80 K, the drastic change appears in the ST-FMR spectra. In addition to the increase in the magnitude of $V_{dc}$, the most apparent difference from $T = 300$ K is its field angular dependence. For example, the non-zero $V_{dc}$ is obtained at $\theta = 0º$ and 180º for $T = 80$ K. It is noted that the single-peak resonance peaks are observed even at $T = 80$ K, where the CSS becomes ferromagnetic. As discussed later in the plot of $H_{Res}$ versus $f$, the ST-FMR signals at $T = 80$ K come from only the CFB layer as well as the result at $T = 300$ K. This is because of large perpendicular magnetic anisotropy (PMA) and the resultant high $H_{Res}$ of ferromagnetic CSS.

**C. Quantitative analysis of spin torque**

In order to quantitatively analyze the spin-torque acting on the CFB layer, the ST-FMR spectra are fitted using the summation of Lorentzian and anti-Lorentzian functions given by $V_{dc} = V_S(\theta)f_L(H) + V_A(\theta)f_{AL}(H)$, respectively, in which $f_L(H) = (\Delta H/2)^2/[(H_{Res} - H)^2 + (\Delta H/2)^2]$ and $f_{AL}(H) = (\Delta H/2)(H_{Res} - H)/[(H_{Res} - H)^2 + (\Delta H/2)^2]$, and $\Delta H$ represents the resonance linewidth. $V_S$ is proportional to the damping-like torque ($\tau_X^0$) whereas $V_A$ is proportional to the field-like torque ($\tau_Y^0$) including the Oersted field contribution [44,45]. **Figure 4(a)** is the spectrum fitted with $f_L(H)$ and $f_{AL}(H)$ measured at $T = 80$ K, $\theta = 0°$ and $f = 16$ GHz. The numerical fitting allows to decompose the spectrum well into the Lorentzian and anti-Lorentzian components (blue and green curves). Consequently, $V_S$ and $V_A$ are evaluated.

The $\theta$ dependence of $V_S$ and $V_A$ for $T = 300$ K is plotted in **Fig. 4(b)**, where $f$ was fixed at 16 GHz. At $T = 300$ K the CSS is paramagnetic, and $V_{dc}$ comes from the oscillating $R(t)$ through the AMR effect of CFB. Here the in-plane angle of magnetization vector of CFB ($\mathbf{M}^{CFB}$) is defined as $\varphi^{CFB}$, and it is assumed that $\mathbf{M}^{CFB}$ follows $\mathbf{H}$, i.e. $\varphi^{CFB} = \theta$. $V_{dc}$ originating from AMR ($V_{dc}^{AMR}$) has the following $\theta$ dependence:

$$V_{dc}^{AMR} \propto \sin 2\theta \left[ f_L(H)\tau_X^0 + \frac{\gamma H_{YY}}{\omega_0} f_{AL}(H)\tau_Y^0 \right], \tag{2}$$

where $\omega_0$ and $\gamma$ are the resonance frequency and the gyromagnetic ratio, respectively. $H_{YY}$ is given by $H_{YY} = H + 4\pi M^{CFB}$. The detailed derivations are given in **Appendix 2**. At the condition of $T = 300$ K, we assume that $\tau_X^0$ corresponds to the torque coming from spin-Hall effect ($\tau_{SHE}$) in the paramagnetic CSS while $\tau_Y^0$ is mostly the current-induced Oersted field torque ($\tau_{Oe}$). $\tau_{SHE}$ and $\tau_{Oe}$ are expressed as

$$\tau_{SHE} = \alpha_{SH} \frac{\gamma \hbar}{2eM_s d^{CFB} d^{CSS}} (I_{rf} \eta_{CSS}) \cos \theta, \tag{3}$$

and

$$\tau_{\text{Oe}} = \frac{\gamma \mu_0}{2w}(I_{\text{rf}} \eta_{\text{Cu+CSS}}) \cos\theta, \tag{4}$$

where $\mu_0$ is the permeability in vacuum, $M_s$ is the saturation magnetization of CFB, and $d^{\text{CFB(CSS)}}$ is the thickness of CFB (CSS) layer. $\eta_{\text{CSS}}$ ($\eta_{\text{Cu+CSS}}$) is the ratio of current flowing in the CSS layer (Cu and CSS layers). From Eqs. (2)-(4), one may expect that $V_S$ and $V_A$ follow the $\sin 2\theta \cos\theta$ dependence. The experimental data in **Fig. 4(b)** are well fitted by the $\sin 2\theta \cos\theta$ function as in the cases of previous study [45].

**Figures 4(c) and 4(d)** show the $\theta$ dependence of $V_S$ and $V_A$ for $T$ = 80 K, where the device was cooled down to 80 K under the application of $H$ = + 5 kOe and – 5 kOe, respectively, along $\theta$ = 45º. The $\theta$ dependence of $V_S$ and $V_A$ for $T$ = 80 K is totally different from that for $T$ = 300 K. Since the CSS is ferromagnetic at $T$ = 80 K, $V_{\text{dc}}$ appears through the AMR effect and/or GMR effect. Based on the experimental fact that the in-plane MR curves exhibit exchange-bias-like behavior (**Fig. 2**), we consider that the CSS layer is divided into three regions: in-plane soft magnetic CSS, in-plane hard magnetic CSS, and out-of-plane hard magnetic CSS as illustrated in **Fig. 2f**. The magnetization vector of in-plane soft magnetic CSS ($\mathbf{M}^{\text{CSS,soft}}$) easily follows $\mathbf{H}$ as well as $\mathbf{M}^{\text{CFB}}$. The MR curve shown in **Fig. 2(c)** suggests $\mathbf{M}^{\text{CSS,soft}}$ and $\mathbf{M}^{\text{CFB}}$ are aligned along $\mathbf{H}$ at $|H|$ > 1500 Oe. On the other hand, the magnetization vector of in-plane hard magnetic CSS ($\mathbf{M}^{\text{CSS,hard}}$) is fixed at $\theta$ = 45º because of the field cooling. The in-plane angles of $\mathbf{M}^{\text{CSS,soft}}$ and $\mathbf{M}^{\text{CSS,hard}}$ are defined as $\varphi^{\text{CSS,soft}}$ and $\varphi^{\text{CSS,hard}}$, respectively. The out-of-plane hard magnetic CSS ($\mathbf{M}^{\text{CSS,OOP}}$) is also not affected by $\mathbf{H}$ because of the strong PMA. In the above situation, we need to consider three processes as possible sources generating $\tau_X^0$: (i) SHE in the ferromagnetic CSS, (ii) SAHE originating from $\mathbf{M}^{\text{CSS,soft}}$ and/or $\mathbf{M}^{\text{CSS,hard}}$, and (iii) spin precession by $\mathbf{M}^{\text{CSS,OOP}}$, which is called the spin precession torque (SPT) [8,9]. The symmetries of $\tau_X^0$ originating from SHE, SAHE, and spin precession are explained in **Appendix 3**. Since $\mathbf{M}^{\text{CSS,soft}}$ and $\mathbf{M}^{\text{CFB}}$ follow $\mathbf{H}$ in the *H* region for the ST-FMR measurement, $\varphi^{\text{CSS,soft}} = \varphi^{\text{CFB}} = \theta$. This case leads to zero $\tau_X^0$ coming from SAHE because it is proportional to $\sin\varphi^{\text{CSS,soft}} \sin(\varphi^{\text{CFB}} - \varphi^{\text{CSS,soft}})$. Therefore, only SAHE from $\mathbf{M}^{\text{CSS,hard}}$ is taken into account for the present experiment. Unfortunately, the current ST-FMR measurement condition does not enable us to separate the contributions from SHE in the ferromagnetic CSS and SAHE originating from $\mathbf{M}^{\text{CSS,hard}}$. In this study, thus, the SHE and the SAHE are evaluated together as a "spin-Hall torque (SHT)". It is noted that $\mathbf{M}^{\text{CSS,OOP}}$ does not contribute to the SAHE because $\mathbf{M}^{\text{CSS,OOP}}$ does not generate the anomalous Hall current along the *z* direction. In addition, the spin precession due to the in-plane magnetized $\mathbf{M}^{\text{CSS,soft}}$ and $\mathbf{M}^{\text{CSS,hard}}$ gives rise to $J_s$ with the out-of-plane-polarized spin, which acts as a field-like torque. Then, the SPT from $\mathbf{M}^{\text{CSS,soft}}$ and $\mathbf{M}^{\text{CSS,hard}}$ is not regarded as a source generating $\tau_X^0$.

To summarize, there are two sources of $\tau_X^0$, *i.e.* SHT ($\tau_{\text{SH}}$) and SPT ($\tau_{\text{SP}}$), and two sources of resistance oscillation, *i.e.* AMR and GMR. These sources lead to the four combinations for the generation process of $V_{\text{dc}}$: SHT+GMR, SPT+GMR, SHT+AMR, and SPT+AMR. $V_{\text{dc}}$ originating from GMR ($V_{\text{dc}}^{\text{GMR}}$) has the following $\theta$ dependence:

$$V_{\text{dc}}^{\text{GMR}} \propto -\sin(\theta - \varphi^{\text{CSS,hard}}) \left[ f_L(H)\tau_X^0 + \frac{\gamma H_{YY}}{\omega_0} f_{AL}(H)\tau_Y^0 \right]. \tag{5}$$

According to the papers reporting on the SPT [8,9], $\tau_{\text{SP}} \propto \sin\theta$. On the other hand, $\tau_{\text{SH}}$ has the same symmetry as that of $\tau_{\text{SHE}}$, namely $\tau_{\text{SH}} \propto \cos\theta$. Then, the $\theta$ dependence of $V_{\text{dc}}$ taking into account both $V_{\text{dc}}^{\text{AMR}}$ and $V_{\text{dc}}^{\text{GMR}}$ is expressed as

$$V_{\text{dc}} = \left[ -V_S^{\text{SHT,GMR}} \sin(\theta - \varphi^{\text{CSS,hard}}) \cos\theta - V_S^{\text{SPT,GMR}} \sin(\theta - \varphi^{\text{CSS,hard}}) \sin\theta + V_S^{\text{SHT,AMR}} \sin 2\theta \cos\theta + V_S^{\text{SPT,AMR}} \sin 2\theta \sin\theta \right] f_L(H) + \left[ -V_A^{\text{Oe,GMR}} \sin(\theta - \varphi^{\text{CSS,hard}}) \cos\theta + V_A^{\text{Oe,AMR}} \sin 2\theta \cos\theta \right] f_{AL}(H). \tag{6}$$

Equation (6) enables us to well fit the $\theta$ dependence of $V_S$ and $V_A$ for $T = 80$ K as shown in **Figs 4(c) and 4(d)**. **Table 1** summarizes the values of $V_S^{SHT,GMR}$, $V_S^{SPT,GMR}$, $V_S^{SHT,AMR}$, $V_S^{SPT,AMR}$, $V_A^{Oe,GMR}$, $V_A^{Oe,AMR}$ and $\varphi^{CSS,hard}$ obtained by the numerical fit for the device cooled with $H = +5$ kOe and $-5$ kOe. The similar values are obtained for each parameter between two field-cooling conditions, *e.g.* $V_S^{SHT,GMR} = 2.30 \pm 0.16$ μV and $2.22 \pm 0.10$ μV for the cooling with $H = +5$ kOe and $-5$ kOe, respectively. On the contrary, $\varphi^{CSS,hard}$ clearly depends on the field-cooling condition, *i.e.* $\varphi^{CSS,hard} \sim 46°$ and $228°$ for the field-cooling conditions with $H = +5$ kOe and $-5$ kOe, respectively. These $\varphi^{CSS,hard}$ values depending on the field-cooling condition are consistent with the results of MR measurement indicating that the $H$ direction during the cooling determines the initial state of $\mathbf{M}^{CSS,hard}$. From the values shown in **Table 1**, the values of $V_S / V_A$ are calculated, which are given in **Table 2**. It should be noted here that the values of $V_S / V_A$ originating from SPT are much smaller than that of SHT, suggesting the small contribution of spin precession process to the spin-torque in the present CSS / Cu / CFB trilayer. In other words, the SHT is the major spin-torque acting on the CFB magnetization. In the next section, we discuss the spin-charge conversion efficiency estimated from these $V_S / V_A$ values.

**D. Enhanced spin-charge conversion efficiency for ferromagnetic CSS**

**Figures 5(a) and 5(b)** plot $f$ versus $H_{Res}$ at $T = 300$ K and 80 K, in which $\theta$ was fixed at 45°. The experimental results obey the Kittel's equation of $f = (\gamma/2\pi)\sqrt{(H_{Res} + H_{ani})(H_{Res} + 4\pi M_{eff})}$, where $H_{ani}$ is the anisotropy field in the film plane and $4\pi M_{eff}$ is the effective demagnetizing field. The numerical fits using the above Kittel's relation give the values of $H_{ani}$ and $M_{eff}$: $H_{ani} = 21$ Oe and $M_{eff} = 689$ emu cm$^{-3}$ for $T = 300$ K and $H_{ani} = 35$ Oe and $M_{eff} = 692$ emu cm$^{-3}$ for $T = 80$ K. These values mean that there is no remarkable temperature dependence in the magnetic properties. As shown in **Fig. 5(c)** representing the *M-T* curve for the 2 nm-thick CFB film, the value of *M* for the CFB layer keeps almost constant in the temperature range below 300 K. Thus, the ST-FMR signal at $T = 80$ K comes from the magnetic resonance of CFB magnetization as well as the result at $T = 300$ K. All the ST-FMR signals in the present experiment are attributable to the CFB magnetization dynamics and are not contaminated with the CSS magnetization even at *T* lower than $T_C$.

**Figure 5(d)** plots the *f* dependence of $V_S / V_A$ obtained from the spectra measured at $T = 80$ K and $\theta = 45°$. $V_{dc}$ sometimes involves the contribution of spin pumping from FM and subsequent inverse SHE in NM [48], and the spin current generated by the spin pump is increased as *f* is increased [49]. One may be aware that $V_S / V_A$ with the negative sign increases slightly with *f*, implying the small contribution of spin pump from CFB and inverse SHE in CSS. As shown in the next paragraph, however, the spin pumping contribution is not so significant and does not affect the evaluation of spin-charge conversion efficiency.

**Figure 5(e)** summarizes the *T* dependence of $V_S / V_A$ measured at $f = 16$ GHz. Here $V_S / V_A$ was evaluated from the fitting to the angular dependence of $V_S$ and $V_A$ as explained in **Section III C**. The orange (green) marks represent the data obtained from the ST-FMR signal originating from AMR (GMR). At $T \leq 150$ K, the ST-FMR signal originating from GMR appears. Near $T_C$, *i.e.* at $T = 150$ K, the GMR contribution to the ST-FMR signal is quite small, giving rise to the remarkable error. With reducing *T* below 100K, the sufficient GMR signal allows the accurate evaluation with the small error, which gives the values being consistent with the values from the ST-FMR signal originating from AMR. There are two important findings. First, the value of $V_S / V_A$ is definitely increased at $T \leq T_C$, suggesting that the phase transition to the ferromagnetic CSS leads to the enhancement of spin charge conversion efficiency. Second, the small GMR signal at $T \sim T_\mathrm{C}$ gives rise to the large error, resulting in the overestimation of $V_\mathrm{S} / V_\mathrm{A}$, and the results from the sufficient AMR signal suggest no particular jump in $V_\mathrm{S} / V_\mathrm{A}$ at $T \sim T_\mathrm{C}$ in the case of CSS. Here, let us discuss again the contribution of spin pumping and subsequent inverse SHE. The field angular dependence of spin pump contribution is the same as that for SHT detected through the AMR. However, the field angular dependence of SHT detected by GMR is totally different from that of spin pumping process. Since the value of $V_\mathrm{S} / V_\mathrm{A}$ from AMR is in agreement with that from GMR at $T \leq 100$ K, we consider that the spin pumping contribution is not so significant.

Then, the spin-charge conversion efficiency ($\xi$) via SHE and/or SAHE is calculated. In order to take into account the contribution of SAHE as well as SHE, $\alpha_\mathrm{SH}$ in Eq. (3) is replaced with $\xi$, and the relationship between $\xi$ and $V_\mathrm{S} / V_\mathrm{A}$ is given by

$$\xi = -\frac{V_S}{V_A}\frac{e\mu_0 M_S^{CFB} d^{CFB} d^{CSS}}{\hbar}\frac{\eta^{CSS+Cu}}{\eta^{CSS}}\sqrt{\frac{H+4\pi M_\mathrm{eff}}{H+H_\mathrm{ani}}}. \qquad (7)$$

As discussed in **Figs. 5(a)-5(c)**, the magnetic properties for CFB do not show the remarkable temperature dependence. In addition, we assume that $\eta^{CSS+Cu}/\eta^{CSS}$ is almost constant against $T$ because of the small monotonic change in $R_\mathrm{xx}$ for the CSS / Cu / CFB trilayer [**Fig. 2(g)**]. Then, we simply consider that $\xi$ is proportional to $V_\mathrm{S} / V_\mathrm{A}$. **Figure 5(f)** plots the $T$ dependence of $\xi$ normalized by the value of $\xi$ at $T = 300$ K, where the values of $V_\mathrm{S} / V_\mathrm{A}$ from the AMR were used. It is noted that $\xi$ at $T = 50$ K reached 140% of that at $T = 300$ K. In our previous work [41], we carefully evaluated $\xi$ to be 0.10 at room temperature employing the similar ST-FMR method. By using this room temperature value, we obtain $\xi = 0.14$ at $T = 50$ K.

**E. Discussion**

In this subsection, first, we again emphasize the importance of taking into account the signal through the GMR effect for analyzing the angular dependence of ST-FMR spectra. Sometime, ST-FMR studies for trilayered structures consisting of FM / NM / FM might overlook the contribution of GMR effect. If the trilayer exhibits a non-negligible GMR effect, one needs to take into account the contribution of ST-FMR signal through the GMR as demonstrated in **Figs. 4(c) and 4(d)**. Otherwise, the angular dependence of ST-FMR signal cannot be analyzed correctly.

Next, the magnitude and the mechanism of spin-charge conversion are discussed. The present CSS shows $\xi = 0.14$ at $T = 50$ K. This $\xi$ is much higher than that for another magnetic Weyl semimetal $Co_2MnGa$ showing $\xi = -0.078$ [25]. Then, we conclude that the ferromagnetic CSS is a material showing highly efficient spin-charge conversion. Although the detailed mechanism for the enhancement is not clear at present, one possible scenario is that the large AHE and the high spin polarization of CSS contribute to the high $\xi$. According to the previous studies on SAHE [6,12,17], the spin anomalous Hall angle, corresponding to the conversion efficiency, is given by the product of the anomalous Hall angle and the spin polarization factors in longitudinal and transverse directions. If this idea is applicable to the present study, since the ferromagnetic CSS possesses the half metallic band structure as well as the huge anomalous Hall angle as reported in [41], the enhancement of $\xi$ may be explained within the framework of SAHE. When the ST-FMR spectra are analyzed in this study, however, the three regions, *i.e.* in-plane soft magnetic CSS, in-plane hard magnetic CSS, and out-of-plane hard magnetic CSS, are regarded as a single source for generating $J_s$ and those contributions are not separated. One may be aware that the SAHE in the out-of-plane hard magnetic CSS do not contribute to $J_s$ flowing in the out-of-plane direction because of the symmetry of AHE. In such a case, we need to examine the $J_s$ generation through the SHE in the out-of-plane hard magnetic CSS with the help from the theoretical calculation, which is given in the next Section. Another point is to exclude the in-plane magnetized CSS that was probably induced during the device fabrication process. This may lead to the further enhancement of spin-charge conversion.

## IV. Theoretical Calculation

### A. Effective tight-binding model of CSS

In this section, we introduce an effective tight-binding model of CSS [50] to theoretically study the intrinsic SHE from CSS. This model reproduces the Weyl points and the nodal line configurations in momentum space, which are similar to those obtained by ab-initio calculations [30,53,54]. In this model, we consider one of $d$ orbitals from Co and $p_z$ orbital from the inter-kagome-layer Sn, which are anticipated to be located near the Fermi level ($E_F$). For simplicity, all other orbitals are neglected. We set primitive translation vectors as $\boldsymbol{a}_1 = (\frac{a}{2}, 0, c)$, $\boldsymbol{a}_2 = (-\frac{a}{4}, \frac{\sqrt{3}a}{4}, c)$, $\boldsymbol{a}_3 = (-\frac{a}{4}, -\frac{\sqrt{3}a}{4}, c)$. In the following we set $c = \frac{\sqrt{3}a}{2}$. The total Hamiltonian is given by,

$$H = H_{\text{d-p}} + H_{\text{so}} + H_{\text{exc}} . \tag{8}$$

Here, $H_{\text{d-p}}$ is the spin independent hopping term,

$$H_{\text{d-p}} = -\sum_{ij\sigma}[t_{ij}d^\dagger_{i\sigma}d_{j\sigma} + t^{\text{dp}}_{ij}(d^\dagger_{i\sigma}p_{j\sigma} + p^\dagger_{i\sigma}d_{j\sigma})] + \epsilon_{\text{p}}p^\dagger_{i\sigma}p_{i\sigma}. \tag{9}$$

$d_{j\sigma}$ and $p_{i\sigma}$ are the annihilation operators of $d$ orbital on Co and $p$ orbital on Sn, respectively. $t_{ij}$ includes the nearest- and second-nearest neighbor hopping, $t_1$ and $t_2$ in the intra kagome layer, and the inter kagome layer $t_z$. $t_{dp}$ is the hybridization between the $d$ orbital and $p$ orbital. $\epsilon_{\text{p}}$ is the on-site potential of the $p$ orbital on Sn.

$H_{\text{so}}$ is a spin-orbit coupling (SOC) term given by $H_{\text{so}} = H^{\text{KM}}_{\text{so}} + H^z_{\text{so}}$. Here, $H^{\text{KM}}_{\text{so}}$ and $H^z_{\text{so}}$ are intra-kagome-layer Kane-Male type SOC [51] and inter-layer-kagome SOC [52], respectively, given

as

$$H_{\text{so}}^{\text{KM}} = -it_{\text{so}}^{\text{KM}} \sum_{\ll ij \gg \sigma\sigma'} v_{ij} d_{i\sigma}^{\dagger} \sigma_{\sigma\sigma'}^{z} d_{j\sigma'}, \tag{10}$$

and

$$H_{\text{so}}^{\text{z}} = -it_{\text{so}}^{\text{z}} \sum_{\ll ij \gg \sigma\sigma'} \boldsymbol{\eta}_{ij} \cdot d_{i\sigma}^{\dagger} \boldsymbol{\sigma}_{\sigma\sigma'} d_{j\sigma'}. \tag{11}$$

In Eq. (10), $t_{\text{so}}^{\text{KM}}$ is the hopping strength and the summation $ij$ is about intra-kagome-layer second nearest-neighbor sites. $v_{ij} = +1(-1)$, when the electron hops counter-clockwise (clockwise) to get to the next-nearest-neighbor site on kagome plane. In Eq. (11), $t_{\text{so}}^{\text{z}}$ is the hopping strength and the summation is about inter-kagome-layer nearest-neighbor hopping. Here, $\boldsymbol{\eta}_{ij}$ is given by $\boldsymbol{\eta}_{CA} = \frac{a_1}{2} \times \frac{a_3}{2} / |\frac{a_1}{2} \times \frac{a_3}{2}|$, $\boldsymbol{\eta}_{AB} = \frac{a_2}{2} \times \frac{a_1}{2} / |\frac{a_2}{2} \times \frac{a_1}{2}|$, $\boldsymbol{\eta}_{BC} = \frac{a_3}{2} \times \frac{a_2}{2} / |\frac{a_3}{2} \times \frac{a_2}{2}|$. Inter-kagome-layer SOC [Eq. (11)] plays an important role to obtain the finite SHC $\sigma_{xz}^{s_y}$.

$H_{\text{exc}}$ is the exchange coupling term between spins of itinerant electrons and magnetization, which is given by

$$H_{\text{exc}} = -J \sum_{i\sigma\sigma'} \boldsymbol{m} \cdot (d_{i\sigma}^{\dagger} \boldsymbol{\sigma}_{\sigma\sigma'} d_{i\sigma'} + p_{i\sigma}^{\dagger} \boldsymbol{\sigma}_{\sigma\sigma'} p_{i\sigma'}). \tag{12}$$

Here, $J$ is the exchange coupling constant and $\boldsymbol{m}$ is the dimensionless magnetization vector. We here consider the exchange coupling on the Sn site for simplicity. In the following, we set $t_1$ as a unit of energy, $t_2 = 0.6t_1$, $t^{\text{dp}} = 1.8t_1$, $t_z = -1.0t_1$, $\epsilon_{\text{p}} = -7.2t_1$, $t_{\text{so}}^{\text{KM}} = -0.2t_1$, $J = 1.2t_1$. These parameters were chosen so that the configurations of the nodal rings are similar to those obtained by ab-initio calculations [53,54]. The chemical potential $\mu$ is determined by using the formula,

$$n_{\text{e}} = \int_{-\infty}^{\infty} d\epsilon \rho(\epsilon) f_{\text{FD}}(\epsilon - \mu, T). \tag{13}$$

Here $n_{\text{e}}$ is the number of the electrons per unit cell and being set as $n_{\text{e}}=3$ in our CSS model, as discussed in Ref. [50,52]. $\rho(\epsilon)$ is the density of states as a function of the energy and $f_{\text{FD}}$ is the Fermi-Dirac distribution function. In the following subsection, the intrinsic SHE is studied based on the effective tight-binding model.

## B. Enhancement of spin Hall conductivity

Then we study the intrinsic SHE with FM ordering by the CSS model. An enhancement of the spin Hall conductivity (SHC) is found in both out-of-plane and in-plane cases as shown in **Figure 6**. We focus on the SHC when the electric field is applied to the $x$-direction and the spin current with $s_y$ flows to the $z$-direction (out-of-plane direction). The SHC characterizing this situation $\sigma_{xz}^{s_y}$ is obtained by the Kubo formula [55],

$$\sigma_{xz}^{s_y} = \frac{e}{4\pi a}\sum_{n\neq m}\int_{BZ} i\frac{d^3k}{(2\pi)^2}\frac{(f_{FD}(E_{nk})-f_{FD}(E_{mk}))}{E_{mk}-E_{nk}} \times \frac{<n\mathbf{k}|v_x|m\mathbf{k}><m\mathbf{k}|j_z^{s_y}|n\mathbf{k}>}{(E_{mk}-E_{nk}+i\eta)}. \quad (14)$$

Here $v_i$ $(i = x, y)$ is the velocity operator given by $v_i = \frac{1}{\hbar}\frac{\partial H(\mathbf{k})}{\partial k_i}$. The spin current operator is given by $j_z^{s_y} = \frac{\hbar}{2}\{v_z, \sigma_y\}$[Ref.2], where $\sigma_y$ is the $y$ component of vector of Pauli matrices. The eigenstates $|n\mathbf{k}>$ are obtained by diagonalizing the total Hamiltonian [Eq. (8)]. **Figure 6** shows the SHC as a function of the amplitude of the magnetic moment using different FM orderings: (a) out-of-plane FM ordering $\mathbf{m} = (0,0,m_z)$ and (b) in-plane magnetic ordering $\mathbf{m} = (0, m_y, 0)$. The system is paramagnetic when $m_z = 0$, whereas the system is fully polarized when $m_z = 1.0$. The SHC is calculated with different strengths of the inter-layer SOC, for $t_{so}^z = -0.0, -0.1t_1$ and $-0.2t_1$. In both cases of out-of-plane and in-plane magnetic orders, it is apparent that the SHC enhances as $m$ increases. These results are consistent with the enhancement of the spin-charge efficiency experimentally observed at $T < T_C$.

## V. Conclusion

The characteristic $T$ dependence of spin-charge conversion for the CSS was found by measuring the ST-FMR for the trilayer consisting of CSS / Cu / CFB. Below $T = 150$ K, where the present CSS layer exhibited the ferromagnetic phase, not only the AMR but also the GMR contributed to the ST-FMR signal. By taking into account the $V_{dc}$ originating from GMR, we successfully explained the field angular dependence of $V_{dc}$ observed at $T < T_C$. We revealed that the SHT involving the torques coming from SHE and/or SAHE plays the major role in the spin torque acting on CFB and the contribution of spin precession process to the spin-torque is negligibly small. A definite increase in $\xi$ was observed at $T < T_C$, indicating that the phase transition to the ferromagnetic CSS leads to the enhancement of spin charge conversion efficiency. The experimental tendency was supported by the theoretical calculation, which showed the increase in spin Hall conductivity with the emergence of magnetic moment at $T < T_C$.

**Acknowledgement**

The authors are grateful to K. Takanashi for valuable discussion and T. Sasaki for her help in the film preparation using the ion beam sputtering. This work was supported by JSPS KAKENHI Grant-in-Aid for Scientific Research (A) (JP20H00299), Grant-in-Aid for Early-Career Scientists (Grant No. JP20K15156), Grant-in-Aid for Scientific Research (B) (JP20H01830), GP-Spin at Tohoku University and JST CREST (JPMJCR18T2).

**Appendix**

**Appendix 1. Temperature dependence of MR curves for coplanar waveguide device**

In addition to the resistance of Hall device, the longitudinal resistance (*R*) of the coplanar waveguide device for the ST-FMR measurement was measured by the two-probe method. **Figure 7** shows MR curves for the coplanar waveguide device measured at (a) *T* = 300 K, (b) 140 K, and (c) 50 K. At *T* = 300 K, A small change in *R* is observed at low *H*, which is attributable to the AMR of CFB layer. As *T* is reduced to 140 K, the large *R* change appears, and the exchange-biased-like behavior is clearly observed in the MR curve measured at *T* = 50 K. *T* dependence of *R* at *H* = 5 kOe ($R^{5kOe}$) and the resistance change ($\Delta R^{MAX}$) is plotted in **Fig. 7(d)**. One sees that the coplanar waveguide devices exhibited the *T* dependence of device resistance similar to those observed for the Hall devices.

**Appendix 2. Derivation of equations for rectification dc voltage**

Let us derive the rectification dc voltage ($V_{dc}$) through the AMR effect of CFB layer and the GMR effect coming from the relative magnetization angle between CFB and CSS layers. As mentioned in the main text, the in-plane angle of magnetic moment of CFB is defined as $\varphi^{CFB}$, and it is assumed that $\mathbf{M}^{CFB}$ follows $\mathbf{H}$ and, *i.e.* $\varphi^{CFB} = \theta$.

In the coordinate depicted in **Fig. 8**, the device resistance (*R*) taking into account the AMR, which depends on the unit vector of CFB magnetization ($\mathbf{m}^{CFB}$), is given by

$$R(\mathbf{m}^{CFB}) = R_0 + \Delta R_{AMR} m_x^2, \qquad (15)$$

where $\mathbf{m}^{CFB} = m_X \mathbf{e}_X + m_Y \mathbf{e}_Y + \mathbf{e}_Z = m_x \mathbf{e}_x + m_y \mathbf{e}_y + m_z \mathbf{e}_z$. $m_Y \mathbf{e}_Y = m_z \mathbf{e}_z$, then $m_X \mathbf{e}_X + \mathbf{e}_Z = m_x \mathbf{e}_x + m_y \mathbf{e}_y$. Since $m_x = \cos\theta - m_X \sin\theta$, taking into account the 1st order term,

$$R(\mathbf{m}^{CFB}) = R_0 + \Delta R_{AMR} \cos^2\theta - \Delta R_{AMR} m_X \sin 2\theta + \cdots \qquad (16)$$

Next, we consider the device resistance change through the GMR effect. Since the experimental MR curves suggest that the GMR comes from the relative angle of magnetization between the CFB and the

Page 21

in-plane magnetized hard magnetic CSS ($\mathbf{m}^{\text{CSS,hard}}$). Then, $R(\mathbf{m}^{\text{CFB}})$ is given by

$$R(\mathbf{m}^{\text{CFB}}) = R_0 - \frac{\Delta R_{\text{GMR}}}{2}(\mathbf{m}^{\text{CFB}} \cdot \mathbf{m}^{\text{CSS,hard}}), \tag{17}$$

where $\mathbf{m}^{\text{CSS,hard}} = m_X' \mathbf{e}_X + m_Z' \mathbf{e}_Z$. The in-plane angle of magnetization of hard magnetic CSS is defined as $\varphi^{\text{CSS,hard}}$, then $\mathbf{m}^{\text{CFB}} \cdot \mathbf{m}^{\text{CSS,hard}} = m_X m_X' + m_Z' = m_X \sin(\theta - \varphi^{\text{CSS,hard}}) + \cos(\theta - \varphi^{\text{CSS,hard}})$. Eq. (17) can be rewritten into

$$R(\mathbf{m}^{\text{CFB}}) = R_0 - \frac{\Delta R_{\text{GMR}}}{2}[m_X \sin(\theta - \varphi^{\text{CSS,hard}}) + \cos(\theta - \varphi^{\text{CSS,hard}})]. \tag{18}$$

Next, we consider the magnetization dynamics under the effective magnetic field ($\mathbf{H}_{\text{eff}}$) and the external torque ($\boldsymbol{\tau}$) is given by the following Landau-Lifshitz-Gilbert equation:

$$\frac{d\mathbf{m}^{\text{CFB}}}{dt} = -\gamma \mu_0 \mathbf{m}^{\text{CFB}} \times \mathbf{H}_{\text{eff}} + \alpha \mathbf{m}^{\text{CFB}} \times \frac{d\mathbf{m}^{\text{CFB}}}{dt} + \boldsymbol{\tau}, \tag{19}$$

where $\boldsymbol{\tau}$ includes spin-transfer torque and current-induced Oersted field torque, which comes from the rf current with the angular frequency of $\omega_p$. $\boldsymbol{\tau}$ is expressed as

$$\tau_X(t) = \tau_X^0 \cos(\omega_p t), \tag{20}$$

$$\tau_Y(t) = \tau_Y^0 \cos(\omega_p t). \tag{21}$$

By linearizing Eq. (19), small deviations of magnetization from the equilibrium points along the X and Y directions are expressed as

$$m_X(t) = C_X \cos(\omega_p t) + D_X \sin(\omega_p t), \tag{22}$$

$$m_Y(t) = C_Y \cos(\omega_p t) + D_Y \sin(\omega_p t), \tag{23}$$

In Eqs. (22) and (23), the phase components same as that of rf current torque of Eqs. (20) and (21) are detected as a rectification voltage. $C_X$ and $C_Y$ are

$$C_X = \frac{1}{\delta} L_S(\omega_p) \tau_X^0 + \frac{1}{\delta} \sqrt{\frac{H_{YY}}{H_{XX}}} L_A(\omega_p) \tau_Y^0, \tag{24}$$

$$C_Y = -\frac{1}{\delta} \sqrt{\frac{H_{YY}}{H_{XX}}} L_A(\omega_p) \tau_X^0 + \frac{1}{\delta} L_S(\omega_p) \tau_Y^0, \tag{25}$$

where $L_{S(A)}$ represents the symmetric (anti-symmetric) Lorentzian function and $\delta$ is the resonance linewidth. $L_S(\omega_p)$ and $L_A(\omega_p)$ are given by

$$L_S(\omega_p) = \frac{(\delta/2)^2}{(\omega_p - \omega_o)^2 + (\delta/2)^2}, \qquad (26)$$

$$L_A(\omega_p) = \frac{(\omega_p - \omega_o)(\delta/2)}{(\omega_p - \omega_o)^2 + (\delta/2)^2}. \qquad (27)$$

Using the relationship of $\delta = (d\omega_0 / dH) \Delta H$, Eqs. (26) and (27) are transformed into

$$L_S(H) = \frac{(\Delta H/2)^2}{(H_{Res} - H)^2 + (\Delta H/2)^2}, \qquad (28)$$

$$L_A(H) = \frac{(H_{Res} - H)(\Delta H/2)}{(H_{Res} - H)^2 + (\Delta H/2)^2}, \qquad (29)$$

where $H_{Res}$ represents the resonance magnetic field.

Here, "torque originating from SHE ($\tau_{SHE}$)", "torque originating from spin precession ($\tau_{SP}$)" for $\tau_X^0$ and "current-induced Oersted field torque ($\tau_{Oe}$)" for $\tau_Y^0$ are considered. The quantization axis of spin generated by the SHE of CSS is along $\mathbf{e}_y = \sin\theta\, \mathbf{e}_Z + \cos\theta\, \mathbf{e}_X$. This means that $\tau_{SHE}$ shows the angular dependence of $\cos\theta$ as shown in Eq. (3). On the other hand, the quantization axis of spin generated by the spin precession by the out-of-plane magnetized CSS ($\mathbf{m}^{CSS,OOP}$) is along $\mathbf{m}^{CSS,OOP} \times (\mathbf{e}_z \times \mathbf{E})$, where $\mathbf{E}$ is the applied electric field. Since $\mathbf{m}^{CSS,OOP} = \mathbf{e}_z$ and $\mathbf{E} = \mathbf{e}_x$, $\mathbf{m}^{CSS,OOP} \times (\mathbf{e}_z \times \mathbf{E}) = \mathbf{e}_x = \cos\theta\, \mathbf{e}_Z - \sin\theta\, \mathbf{e}_X$. This means that $\tau_{SP}$ shows the angular dependence of $\sin\theta$. Then, $\tau_{SP}$ is expressed as

$$\tau_{SP} = -\alpha_{SP} \frac{\gamma\hbar}{2eM_s d^{CFB} d^{CSS}} (I_{rf}\eta_{CSS}) \sin\theta, \qquad (30)$$

where $\alpha_{SP}$ represents the spin-charge conversion efficiency through the spin precession process. $\tau_{Oe}$ is given by $\tau_{Oe} = -\gamma\mu_0 \mathbf{m}^{CFB} \times \mathbf{H}_{Oe}$, where $\mathbf{H}_{Oe}$ is the Oersted field along $-\mathbf{e}_y$ and given by $\mathbf{H}_{Oe} = -(I_{rf}\eta_{Cu+CSS}/2w)\mathbf{e}_y$. Then, we obtained Eq. (4). In summary, since $\tau_X^0 = \tau_{SHE} + \tau_{SP}$ and $\tau_Y^0 = \tau_{Oe}$,

$$\tau_X^0 = \frac{\gamma \hbar I_{rf} \eta_{CSS}}{2eM_s d^{CFB} d^{CSS}} (\alpha_{SH} \cos\theta - \alpha_{SP} \sin\theta), \tag{31}$$

$$\tau_Y^0 = \mu_0 \frac{\gamma I_{rf} \eta_{Cu+CSS}}{2w} \cos\theta. \tag{32}$$

Finally, $V_{dc}$ through AMR and GMR effect is obtained. From Eqs. (16) and (18), the resistance change depending on $m_X, m_Y$ ($\Delta R(m_X, m_Y)$) is given by

$$\Delta R(m_X, m_Y) = \Delta R_{AMR} m_X \sin 2\theta - \frac{\Delta R_{GMR}}{2} m_X \sin(\theta - \varphi^{CSS,hard}). \tag{33}$$

Then, the multiplication of $\Delta R(m_X, m_Y)$ and $I_{rf}$ leads to the time-independent voltage, which is

$$V_{dc} = \langle \Delta R(m_X, m_Y) \cdot I_{rf} \rangle$$
$$= \frac{1}{2} \Delta R_{AMR} C_X I \sin 2\theta - \frac{1}{4} \Delta R_{GMR} C_X I \sin(\theta - \varphi^{CSS,hard}), \tag{34}$$

where

$$C_X = \frac{1}{\delta} L_S(\omega_p) \left[ \frac{\gamma \hbar I_{rf} \eta_{CSS}}{2eM_s d^{CFB} d^{CSS}} (\alpha_{SH} \cos\theta - \alpha_{SP} \sin\theta) \right]$$
$$+ \frac{1}{\delta} \sqrt{\frac{H_{YY}}{H_{XX}}} L_A(\omega_p) \left[ \mu_0 \frac{\gamma I_{rf} \eta_{Cu+CSS}}{2w} \cos\theta \right]. \tag{35}$$

**Appendix 3. Symmetries of $\tau_X^0$ originating from SHE, SAHE, and spin precession**

In the case of **E** // **x**, the SHE of CSS generates $J_s$ flowing along the $z$ direction with the quantization axis of spin along $\mathbf{e}_y$. Thus, the $\tau_X^0$ coming from the SHE has the symmetry of $\mathbf{m}^{CFB} \times [(\mathbf{e}_z \times \mathbf{E}) \times \mathbf{m}^{CFB}]$.

Next, let us consider the symmetry of SAHE related with the magnetic moment for in-plane hard magnetic CSS. $J_s$ due to the SAHE is proportional to $\mathbf{m}^{CSS,hard} \times \mathbf{E}$. Then, the $\tau_X^0$ coming from the SAHE has the symmetry of $(\mathbf{m}^{CSS,hard} \times \mathbf{E}) \cdot \mathbf{e}_z [(\mathbf{m}^{CFB} \times \mathbf{m}^{CSS,hard}) \times \mathbf{m}^{CFB}]$.

As described in **Appendix 2**, the quantization axis of spin generated by the spin precession by the out-of-plane magnetized CSS is along $\mathbf{m}^{CSS,OOP} \times (\mathbf{e}_z \times \mathbf{E}) = \mathbf{m}^{CSS,OOP} \times \mathbf{e}_y$. Then, the $\tau_X^0$ coming from the spin precession by the out-of-plane magnetized CSS has the symmetry of $\mathbf{m}^{\mathrm{CFB}} \times [(\mathbf{m}^{\mathrm{CSS,OOP}} \times \mathbf{e}_y) \times \mathbf{m}^{\mathrm{CFB}}]$.

**Table 1** Estimated values of $V_S^{SHT,GMR}$, $V_S^{SPT,GMR}$, $V_S^{SHT,AMR}$, $V_S^{SPT,AMR}$, $V_A^{Oe,GMR}$, $V_A^{Oe,AMR}$ and $\varphi^{CSS,hard}$ by numerical fit using Eq. (6) to in-plane field angular $\theta$ dependence of $V_S$ and $V_A$ for $T = 80$ K, in which the device was cooled with $H = +5$ kOe and $-5$ kOe along $\theta = 45°$.

|  | $V_S^{SHT,GMR}$ ($\mu$V) | $V_S^{SPT,GMR}$ ($\mu$V) | $V_S^{SHT,AMR}$ ($\mu$V) | $V_S^{SPT,AMR}$ ($\mu$V) | $V_A^{Oe,GMR}$ ($\mu$V) | $V_A^{Oe,AMR}$ ($\mu$V) | $\varphi^{CSS,hard}$ (°) |
|---|---|---|---|---|---|---|---|
| Cooling $H = +5$ kOe | 2.30 ± 0.16 | -0.01 ± 0.21 | 1.11 ± 0.18 | -0.26 ± 0.17 | -3.52 ± 0.11 | -1.83 ± 0.08 | 46 ± 3 |
| Cooling $H = -5$ kOe | 2.22 ± 0.10 | 0.05 ± 0.13 | 1.11 ± 0.12 | 0.24 ± 0.11 | -3.92 ± 0.23 | -2.06 ± 0.17 | 228 ± 2 |

**Table 2** Calculated values of $V_S / V_A$ for the processes of SHT+GMR, SPT+GMR, SHT+AMR, and SPT+AMR at $T = 80$ K, in which the device was cooled with $H = +5$ kOe and $-5$ kOe along $\theta = 45°$.

|  | SHT+GMR | SPT+GMR | SHT+AMR | SPT+AMR |
|---|---|---|---|---|
| Cooling $H = +5$ kOe | -0.65 | 0 | -0.61 | 0.14 |
| Cooling $H = -5$ kOe | -0.57 | -0.01 | -0.54 | -0.12 |

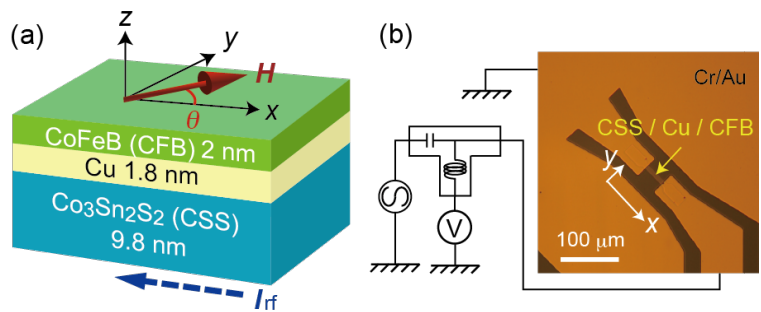

**Figure 1** (single column)

(a) Schematic illustration of thin film structure, and (b) optical microscope image of coplanar waveguide (CPW) device together with the measurement circuit for spin-torque ferromagnetic resonance (ST-FMR). The magnetic field ($H$) was applied in the film plane with the angle of $\theta$, and the rf current ($I_{rf}$) was applied along the $x$ direction.

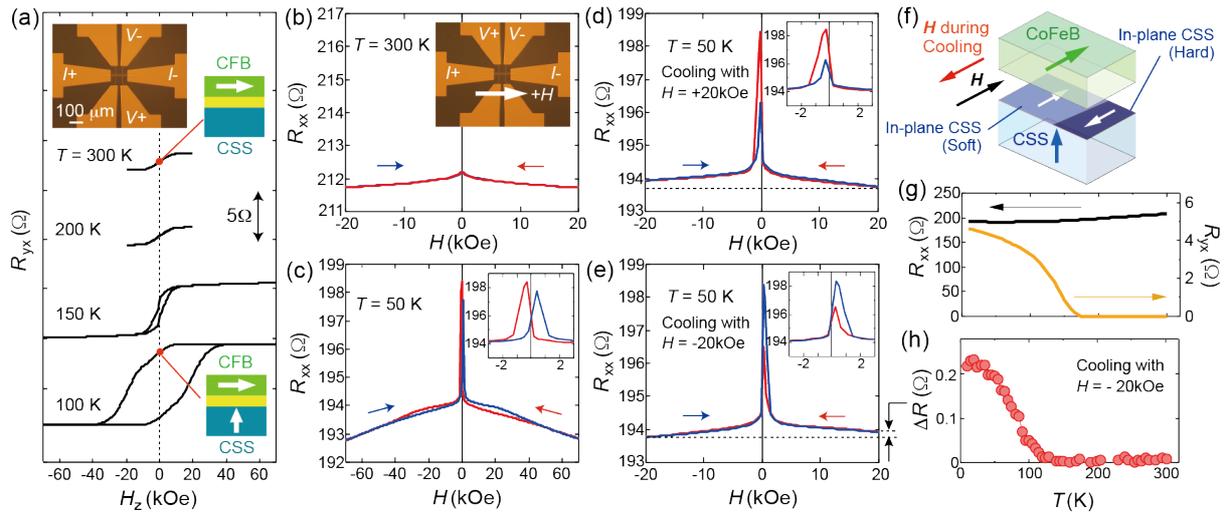

**Figure 2** (double column)

(a) Transverse resistance ($R_{yx}$) versus out-of-plane magnetic field ($H_z$) for the Hall device measured at $T$ = 300 K, 200 K, 150 K, and 100 K. The insets display the optical microscope image with current and voltage probes and the schematic illustrations of magnetization configurations. The $R_{yx}$ - $H_z$ curves were vertically shifted for clarity. (b) Magnetoresistance (MR) curve measured at $T$ = 300 K. The longitudinal resistance ($R_{xx}$) was measured using the four-probe method, and the in-plane magnetic field ($H$) was applied along the channel of Hall device as shown in the inset. (c) Full MR curve measured at 50 K with $H$ applied in the range of ± 70 kOe, and (d) minor MR curve with $H$ applied in the range of ± 20 kOe. For both measurements, the device was cooled down to 50 K under the application of $H$ = + 20 kOe. (e) Minor MR curve, in which the device was cooled down with $H$ = – 20 kOe applied. The red (blue) arrow denotes the field sweep direction from positive (negative) to negative (positive). In (c) (d) and (e), the MR curves enlarged at low $H$ regions are shown as the insets. The black dotted lines in (d) and (e) compare the values of $R_{xx}$ at + 20 kOe and – 20 kOe, indicating that there exists the difference in $R_{xx}$ between $H$ = + 20 kOe and – 20 kOe, which was defined as $\Delta R$. (f) Schematic illustration of possible magnetic structures in CFB and CSS when $H$ (denoted by black arrow) was applied opposite to the magnetic field during cooling the device (denoted by red arrow), which corresponds to the magnetization configuration in the minor MR curve. (g) $R_{xx}$ and $R_{yx}$ as a function of $T$. (h) $\Delta R$ as a function of $T$, in which the device was cooled down with $H$ = – 20 kOe applied.

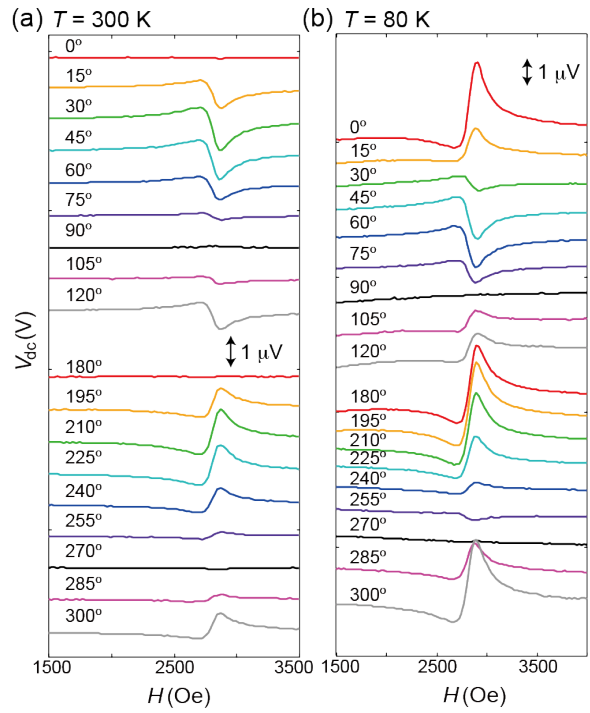

**Figure 3** (single column)

ST-FMR spectra measured at (a) $T$ = 300 K and (b) 80 K, in which the excitation frequency ($f$) was fixed at 16 GHz while $\theta$ was varied. For clarity, the spectra were shifted vertically.

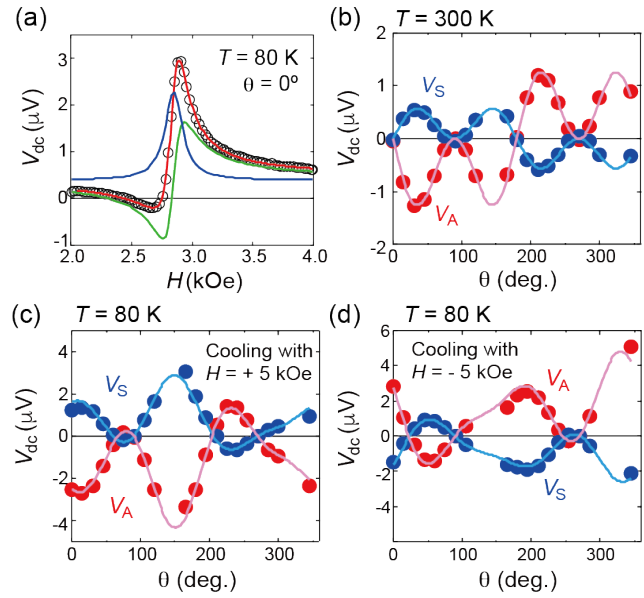

**Figure 4** (Single column)

(a) Fitting curves for the ST-FMR spectrum obtained at $T = 80$ K, $\theta = 0°$ and $f = 16$ GHz, where $T$ was reduced under the application of $H = +5$ kOe along $\theta = 45°$. (b) $\theta$ dependence of detected dc voltage ($V_{dc}$) measured at $T = 300$ K and $f = 16$ GHz. (c)-(d) $\theta$ dependence of $V_{dc}$ measured at 80 K and $f = 16$ GHz, in which $T$ was reduced under the application of (c) $H = +5$ kOe and (d) $-5$ kOe along $\theta = 45°$. The blue and red marks represent the experimental values of symmetric ($V_S$) and antisymmetric components ($V_A$), respectively, and the solid curves represent the fitting results.

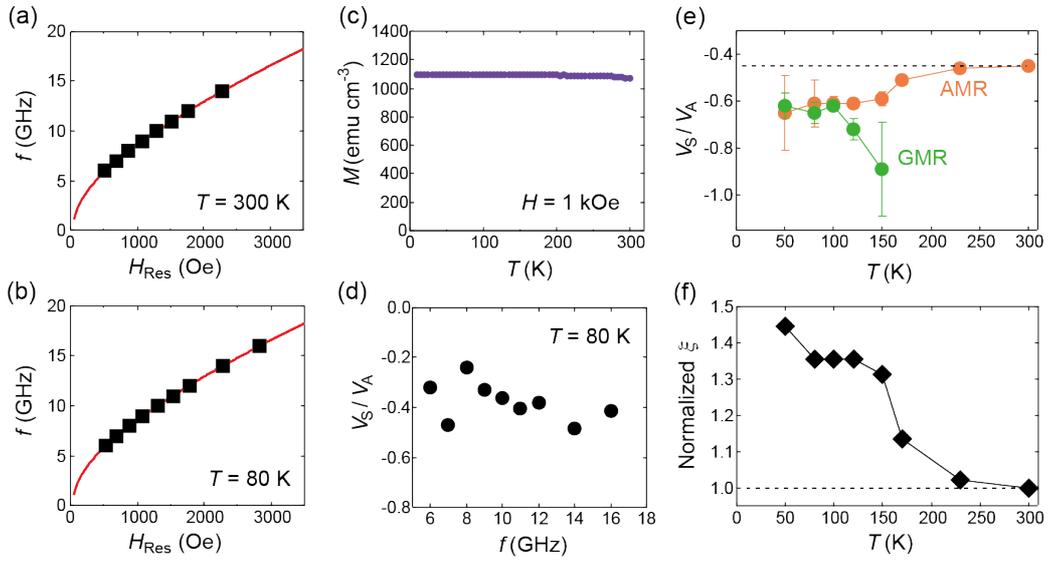

**Figure 5** (Double column)

(a) Resonant magnetic field ($H_{Res}$) versus $f$ at $T$ = 300 K and (b) 80 K, where $\theta$ was fixed at 45°. The solid squares denote the experimental data whereas the solid lines denote the results of fitting. (c) $T$ dependence of magnetization ($M$) for the 2 nm-thick CFB film measured at $H$ = 1 kOe. (d) $f$ dependence of $V_S / V_A$, where $\theta$ = 45° and $T$ = 80 K. (e) $T$ dependence of $V_S / V_A$ obtained from the AMR (orange marks) and the GMR signals (green marks) measured at $f$ = 16 GHz. (f) $T$ dependence of spin-charge conversion efficiency ($\xi$) normalized by the value at $T$ = 300 K.

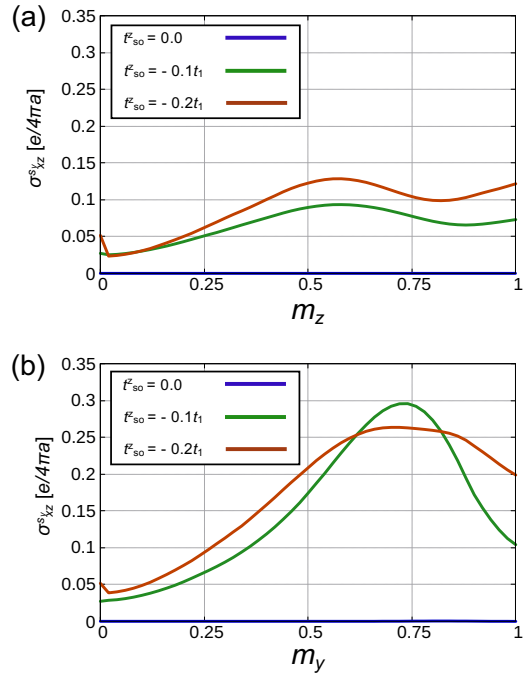

**Figure 6** (Single column)

Magnetic moment amplitude dependence of the spin Hall conductivity for the CSS model with (a) out-of-plane magnetization and (b) in-plane magnetization for $t^z_{\text{so}} = 0.0, -0.1t_1$ and $t^z_{\text{so}} = -0.2t_1$.

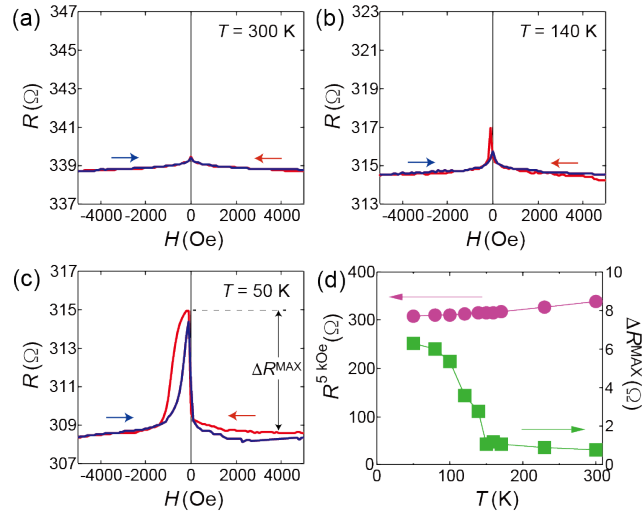

**Figure 7** (Single column)

MR curves for the coplanar waveguide device measured at (a) $T$ = 300 K, (b) 140 K, and (c) 50 K. The resistance for the coplanar waveguide device ($R$) was measured using the two-probe method, and the in-plane $H$ was applied at $\theta$ = 45°. (d) $R$ at $H$ = 5 kOe ($R^{5kOe}$) and the resistance change ($\Delta R^{MAX}$), which is defined in (c), as a function of $T$.

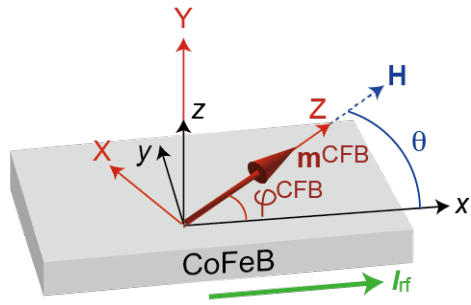

**Figure 8** (Single column)

Schematic illustration of two coordinate systems: ($x$, $y$, $z$) and ($X$, $Y$, $Z$). In the equilibrium condition, the magnetic moment of CFB ($\mathbf{m}^{CFB}$) is parallel to $\mathbf{e}_Z$, which is also parallel to $\mathbf{H}$, *i.e.* $\varphi^{CFB} = \theta$. $I_{rf}$ was applied along $\mathbf{e}_x$.